\documentclass[11pt,a4paper]{article}
\usepackage{jheppub}
\usepackage{fancybox}
\usepackage{color}
\usepackage{pstricks}

\usepackage{young}
\usepackage{color}
\usepackage{pstricks}
\usepackage{graphicx}

\title{Holographic Fermionic Fixed Points in d=3}
\author[a]{Joshua L. Davis, Hamid Omid}
\author[a,b]{Gordon W. Semenoff}

\affiliation[a]{Department of Physics and Astronomy, University of British Columbia,\\
Vancouver, British Columbia, Canada V6T 1Z1}
\affiliation[b]{Kavli Institute for Theoretical Physics, University of California, Santa Barbara, California 93106-4030\\
Preprint:
{\bf NSF-KITP-11-135}}
\emailAdd{jdavis@phas.ubc.ca}
\emailAdd{omidh@phas.ubc.ca}
\emailAdd{gordonws@phas.ubc.ca}

\abstract{We present a top-down string theory holographic model of
strongly interacting relativistic $2+1$-dimensional fermions, paying
careful attention to the discrete symmetries of parity and time reversal invariance. Our construction is based on probe $D7$-branes in $AdS_5 \times S^5$, stabilized by internal fluxes.
We find three solutions, a parity and time reversal invariant conformal field theory which can
be viewed as a
particular deformation of Coulomb interacting graphene, a parity and time reversal violating but gapless
field theory and a system with a parity and time reversal violating charge gap.
We show that the Chern-Simons-like electric response function, which is generated perturbatively
at one-loop order by parity violating fermions and which is protected by a no-renormalization theorem
at orders beyond one loop, indeed appears with the correctly quantized coefficient in the charge gapped theory.
In the gapless parity violating solution, the Chern-Simons response function obtains quantum corrections
which we compute in the holographic theory.
}

\begin{document}
\maketitle

\section{Introduction and Summary}

 The $AdS/CFT$ correspondence \cite{maldacena} has recently been used extensively
 as a tool to examine the behavior of quantum systems at strong coupling \cite{hartnoll}.
Emergent relativistic 2+1-dimensional fermions in condensed matter systems, such as graphene \cite{Semenoff:1984dq}
\cite{graphene},
topological insulators \cite{kane},
the D-wave state of high $T_c$ superconductors \cite{balents} and optical
 lattices \cite{dunne} lead to interesting
 dynamical questions, often in the strong coupling regime.

 For example, a model of the low energy ($<1ev$) physics of graphene
 has massless relativistic fermions interacting with a Coulomb potential and emergent $SU(4)$ symmetry,
 \begin{align}
 S=&\int d^3x~\sum_{k=1}^4 \bar\psi_k\left[\gamma^t(i\partial_t-A_t)+v_F\vec\gamma\cdot(i\vec\nabla-\vec A)\right]\psi_k
  -\frac{1}{4e^2 }\int d^3x~F_{ab}\frac{1}{2\sqrt{-\partial^2}}F^{ab}
 \label{graphene}
 \end{align}
 where we have set $\hbar=1=c$. Electron dynamics are confined to the 2-dimensional sheet of carbon atoms
 which composes graphene.  The nonlocal gauge field action arises from
 the fact that the photon propagates in the 3-dimensional space surrounding the graphene sheet,
 This model does not have $SO(2,1)$ Lorentz symmetry since the speeds of the
 massless electron and the photon are not identical, in fact, $v_F\sim c/300$.
 To a first approximation, the Coulomb interaction is instantaneous and the
 magnetic interactions are suppressed by factors of $\tfrac{v_F}{c}$.

 The field theory with action (\ref{graphene}) is scale invariant at the classical level and
  it is known that, at least up to two loop order, it is a renormalizable
 quantum field theory in the sense that no counterterms with new interactions have to be introduced to cancel
 ultraviolet divergences \cite{voz}.  Upon renormalization, the graphene speed of light $v_F$ requires a logarithmically divergent
 counterterm and it therefore becomes a scale-dependent running coupling constant which increases to larger values
 in the infrared limit. This running should be cutoff when $v_F$ reaches the speed of light.  The resulting Lorentz
 invariant theory is conjectured to be a conformal field theory for any value of the dimensionless coupling constant, $e$.   This is confirmed to two loop order in perturbation theory.

 However, to describe realistic physics of graphene, (\ref{graphene}) is nominally a strongly interacting model.
 The strength of Coulomb interactions
  are governed by the graphene fine structure constant, which
 (if we use the vacuum dielectric constant $\epsilon_0=1$) is
 $$
 \alpha_g= \frac{e^2}{4\pi\hbar v_F}\approx \tfrac{300}{137.035...} > 1
 $$
 Notwithstanding the fact that, if $v_F$ runs to larger values, $\alpha_g$ runs to smaller values in the
 infrared, if the fine structure constant is really this large, the accuracy of perturbative computations is
 questionable at best.

In this paper, we shall formulate a top-down, first principles construction of a class of models containing 2+1-dimensional relativistic fermions using the $AdS/CFT$ correspondence.  We believe that these models are as close as one can come with
the current state of the field to a holographic model of graphene.  Moreover, they should have wider applicability, particularly
to topological insulators which can also be strongly interacting systems.
We will discuss their field theory dual in more detail
shortly.  For now, we note that they and the graphene model (\ref{graphene}) have some significant differences.
For example, the construction that we use has $SO(2,1)$ Lorentz invariance and can therefore only describe
(\ref{graphene}) when $v_F=c$.  In addition, (\ref{graphene}) has a $U(1)$ gauge symmetry. Our holographic
model generalizes it to $U(N)$ gauge symmetry, and then analyzes it in the limit $N\to\infty$, a long way from $N=1$.
This is not as bad as it sounds.  As we will discuss in later sections, if one combines the charges of particles $g_{\rm YM}$
with factors of $N$, in the planar limit, $N$ itself will make an appearance only through the 't
Hooft coupling, $\lambda=g_{\rm YM}^2N$
which is held fixed in that limit.  The holographic construction is accurate when this coupling is large and
in that sense, the large values of planar gauge theory fine structure constant,
$\frac{\lambda}{4\pi\hbar c}$ could emulate the large fine structure constant $\alpha_g$ of graphene.
Furthermore, the system which we analyze does have gluon-mediated current-current interactions
with a $\frac{1}{x^2}$ propagator very similar
in nature to the electromagnetic interaction in (\ref{graphene}).
The planar limit of the theory does take this into account
 to all orders in the coupling constant, and the coupling is then taken to be large.  The main differences are then
 the addition of higher correlations of the gluon (the three and four gluon vertices) and the truncation of
 Feynman diagrams to those which are planar, two effects which compensate each other somewhat.

 Of course there is the question as to whether, if the coupling becomes strong enough, a system such as graphene
 will generate a mass gap for the fermions by spontaneously breaking the flavor symmetry ($U(4)$ in the graphene
 model (\ref{graphene})).
 There is at present no experimental evidence for such a gap, at least in
 the absence of strong magnetic fields.   In this paper we will find a strong coupling limit where the fermion
 spectrum does not
 acquire a gap, but where the strong coupling limit is a nontrivial conformal field theory.
 The different construction, where it does obtain a gap has been examined in references \cite{rey} and
 \cite{kp}. Our solution at least demonstrates that there is a $D$-brane construction of a strongly interacting theory where
 gap generation does not occur. In fact, it turns out to be very difficult to find a gapped solution in the class of models
 that we consider. We do find one, but it is  only in a system which has extra degrees of freedom.

A general quantum field theory of the form we want to emulate
is described by the action
%
%
\begin{align}\label{fermionlagrangian}
S =\int d^3x~\bar\psi_k(i\gamma^a\partial_a+im)\psi_k  +~{\rm interactions}
\end{align}
Here, and in the rest of this paper, we assume that 2+1-dimensional spacetime has Euclidean signature.  (Dirac gamma matrices
are assumed to be hermitian and the Euclidean space mass term has a factor of $i$.)
The theory described by (\ref{fermionlagrangian})
contains $N$ species ($k$ is summed from
1 to N) of a 2-component spinor field $\psi_k$ which
are complex.  As well as gapping the fermion spectrum, the mass term violates parity and time reversal invariance.
It is
absent in the graphene model (\ref{graphene}) which is parity invariant, but could be present in a time reversal violating
topological insulator, for example.

\begin{figure}
 ~~~~~~~~~~~~~~~~~~~~~~~~~~~~~~~~~~~~~~~~~~~~~~~~~\includegraphics[scale=.8]{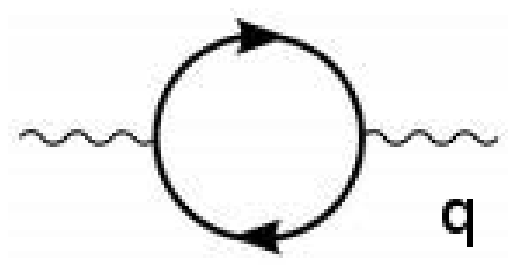}\\
\begin{caption} {The Feynman diagram with is computed to find the result quoted in eq.~(\ref{vacuumpolarization1}).  \label{vacuumpolarization}
}\end{caption}
\label{dbranes}\end{figure}

The model (\ref{fermionlagrangian}) has a conserved
$U(1)$ charge with current
\begin{equation}
j_a(x)=\sum_{k=1}^N\bar\psi_k(x)\gamma_a\psi_k(x)
\end{equation}
An interesting probe of the properties of this field theory is the current-current correlation function,
\begin{equation}\label{currentcurrent}
\left<j_a(x)j_b (0)\right>=\int \frac{d^3q~e^{iqx}}{(2\pi)^3}\Delta_{ab}(q)\end{equation}
Lorentz invariance implies that it must have the form
\begin{align}\label{response}
\Delta_{ab}(q) = \Delta_{\rm CS}(q)~\epsilon_{abc}q^c
+\Delta_T(q) (q^2 \delta_{ab}- q_aq_b )
\end{align}
with two functions $\Delta_{\rm CS}(q)$ and $\Delta_{\rm T}(q)$ of the magnitude of $q$.
For $N$ two-component Dirac spinors
with a mass, $m$, as in (\ref{fermionlagrangian}), and at order one loop in
perturbation theory, the Feynman diagram in figure \ref{vacuumpolarization} yields
\begin{align}\label{vacuumpolarization1}
\Delta_{\rm CS}(q)&=\frac{N}{2\pi}\frac{m}{  q}\arctan\frac{q}{2m} \\
\Delta_{\rm T}(q)&=\frac{N}{4\pi}\left[\frac{m}{q^2}+\frac{1}{2q}(1-\tfrac{4m^2}{q^2})\arctan\frac{q}{2m}\right]
\end{align}
These functions have the large  and small momentum limits
\begin{align}
q>>m:&~~~\Delta_{\rm CS}(q)=\frac{N}{4 }\frac{m}{q}+\ldots~~,~~
\Delta_T(q)=\frac{N}{16 q}+\ldots
\\
q<<m:&~~~\Delta_{\rm CS}(q)=\frac{N}{4\pi}\frac{m}{|m|}+\ldots~~,~~
\Delta_T(q)=\frac{N}{12\pi|m|}+\ldots
\label{massiveresponse}
\end{align}
respectively.
The Fermion mass in (\ref{fermionlagrangian}) violates parity and time reversal
invariance explicitly and the parity and time reversal violating
first term in (\ref{response}) is a result.
The leading, zero momentum contribution to
$\Delta_{\rm CS}$ in (\ref{massiveresponse}) is independent of the  magnitude of
the mass $m$ and only depends on its sign.
This is a manifestation of the parity anomaly \cite{Niemi} and leads to an induced
Chern Simons term in the effective action of a gauge field which would couple to the $U(1)$ current.
It also results in a quantum Hall effect, even in the absence of magnetic field, with a
half-quantized Hall conductivity
\begin{equation}\sigma_{xy}~=~\frac{1}{2}~\frac{e^2N}{2\pi\hbar}\end{equation}
Moreover, $\Delta_{\rm CS}(q=0)$
has a no-renormalization theorem \cite{Coleman:1985zi},
for a large class of interactions which preserve the charge gap\footnote{Here, charge gap refers to the existence of a mass
gap for all excitations of the theory which carry electric charge.  Ordinary $3+1$-dimensional quantum electrodynamics has no mass gap, since the photon is massless, but it does have a charge gap, since the electron has mass.  The existence of a charge gap is
sufficient to make the current-current correlations functions analytic in the region where the 4-momenta carried
by the currents approach zero.}, either relativistic or nonrelativistic,
corrections vanish order by order in perturbation theory beyond one loop.

On the other hand, it has been demonstrated explicitly
that, if the charge gap vanishes, {\it i.e.} if there are any massless charged particles in the spectrum,
$\Delta_{\rm CS}(0)$ can renormalize at two and higher loop orders \cite{semenoffsodanowu}. Then, $\Delta_{\rm CS}(0)$ can differ
from that in (\ref{massiveresponse}).  In addition, $\Delta_T(q)$ should obtain the cut singularity $\Delta_{\rm T}\sim {1}/{\sqrt{q^2}}$  that is expected
for massless charged particles. The latter is indeed seen at one loop order if one puts $m=0$
in  (\ref{fermionlagrangian}), $\Delta_{\rm T}(q) = \frac{N}{16q} $. Both the renormalization of $\Delta_{\rm CS}(q=0)$ and the nonanalyticity $\Delta_{\rm T}\sim\tfrac{1}{q}$ are diagnostics of the presence of charged massless degrees
of freedom.  In particular, if a parity and time reversal violating conformal field theory existed, conformal invariance
and the fact that $j_a$ is conserved determine the current-current correlator up to constants, where  $\Delta_{\rm CS}={\rm constant}$ and
$\Delta_{\rm T}= {\rm constant}/q $ would be exact statements for all momenta.
If it were parity or time reversal invariant, then $\Delta_{\rm CS}(q)$ would vanish.


In this paper, we shall find holographic realizations of  three
scenarios.  The first is a parity and time reversal invariant conformal field theory which can be regarded
 as large N deformation of the graphene model (\ref{graphene}) with  $v_F=c$.  Secondly, we shall find
 a parity and time reversal
violating but still gapless theory which flows to a parity and time reversal violating conformal field theory at
an infrared fixed point.  Thirdly, we will find a  theory with a charge gap
whose current-current correlator
has functions resembling  (\ref{massiveresponse}) with the correctly
quantized  $\Delta_{\rm CS}(0)=\tfrac{N}{4\pi}$, consistent with the
perturbative no-renormalization theorem. The latter is important.  It can be regarded as a nontrivial check of the construction -- a match between weak and strong coupling of an (albeit coupling constant independent) quantity.  All three of our constructions correspond to field theories which at high energies approach the parity and time reversal invariant conformal field theory.
The  parity violating conformal field theory
we find as the infrared limit of our second solution can be regarded as
the strong coupling limit of a model of the quantum Hall plateau
which uses Dirac fermions coupled to a $U(1)$ Chern-Simons gauge theory \cite{chenfisherwu}.
Ref.\ \cite{chenfisherwu} demonstrated that such an infrared fixed point exists
at weak coupling up to order two loops.  Our result implies that
such a theory can also exist at strong coupling.

\begin{figure}
 ~\includegraphics[scale=.6]{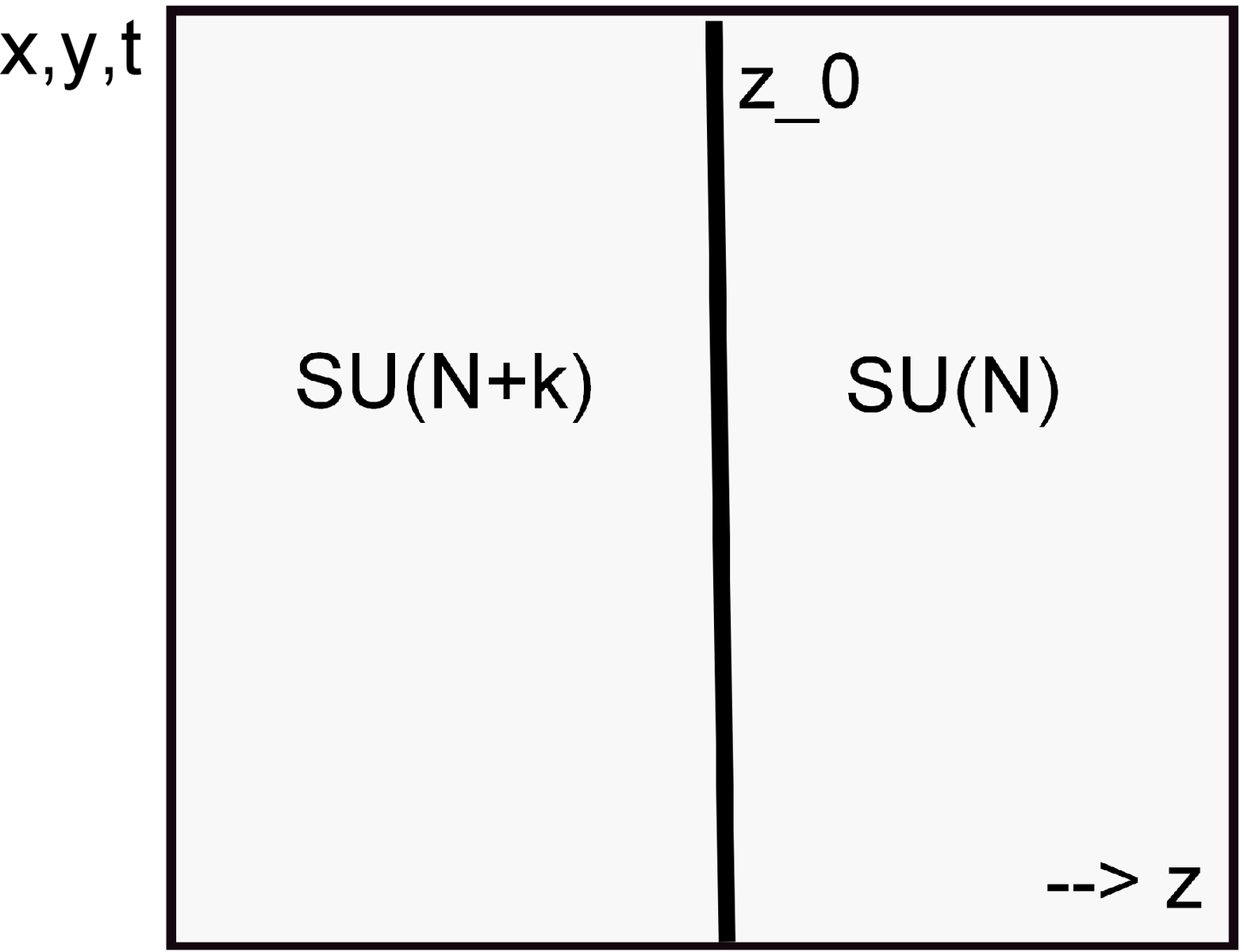}\\
\begin{caption} {We are studying a defect conformal field theory were fermions are constrained to occupy
a plane, denoted by the vertical line through the center of the diagram.   This plane divides the three dimensional
space into two different regions which are occupied by four dimensional conformal ${\cal N}=4$ supersymmetric
Yang-Mills theories with different gauge groups.  The conformal field theory has three tunable parameters, $N$, $N+k$
and the Yang-Mills coupling constants.  The holographic description describes the planar limit of this theory where
the 't Hooft coupling is tuned to be large.   The remaining parameter is $k=n_D^2$.  The
anomalous dimensions of operators and the current-current correlation functions will turn out to be dependent on this
parameter (through $f$ in the following sections). \label{cft}
}\end{caption}
\end{figure}

The field theory dual of the system that we shall study is a defect conformal field theory consisting
of fermions confined to move on a 2-dimensional plane which separates 3-dimensional space into two regions
as depicted in figure \ref{cft}.
The fermions carry a global $U(1)$ charge which couples to the current $j_a(x)$ whose correlation functions
we discussed in eq.~(\ref{currentcurrent}).  They also carry the fundamental representation of the
gauge group of  ${\cal N}=4$ supersymmetric Yang-Mills theory
and they interact by exchanging gluons and other particles in that gauge theory.  The interaction-mediating
degrees of freedom are allowed to propagate in the 3+1-dimensional space-time whereas the fermions themselves are
constrained to occupy the 2+1-dimensional space-time.  The holographic model that we construct has Yang-Mills
theory with different gauge groups on each side of the defect, as shown in figure \ref{cft}.
On the D-brane side, this
is a result of the fact that, the D7-branes with internal fluxes which we shall use are technically D7-branes
with $n_D^2$ D3-branes dissolved into their worldvolumes.  The D7-brane then forms a boundary between regions with
different numbers of D3-branes and therefore different amounts of RR flux, thus different ranks of the gauge group in the
field theory dual.  We shall describe the details of our top-down construction in the next section.

The ${\cal N}=4$
supersymmetric Yang-Mills theory is a four dimensional conformal field theory.  In particular, this implies that
its global color current-current interaction varies as $\frac{1}{4\pi x^2}$, there $x$ is the space-time interval,
identical to the current-current interaction is 3+1-dimensional electrodynamics. This feature persists in
the holographic construction, at strong coupling.   One of the mysteries of graphene is the fact that, even though
the interactions in the model (\ref{graphene}) are strong, much of the phenomenology is modeled very well by free
fermions.  We shall see that some properties of the conformal field theory, like electromagnetic response, that we
find in the strong coupling limit bear a strong resemblance to what one would find for free fermions.  Others,
such as the anomalous dimension of the mass operator do not, they are tuneable and could differ significantly from
the their weak coupling limits.  As an example of a quantity which is similar, conformal invariance fixes the
AC conductivity which is a nonzero constant even in the charge neutral state of the theory where the electron density
vanishes.  This behavior is seen in graphene and it has been analyzed in the weak coupling regime.  Here we see
that it is modified very little in the strong coupling regime of our model.  Experimental estimates of this neutral
point conductivity vary but are typically of the same order as the results that we find.

\section{Holography}

The holographic construction that we shall pursue has become
possible after the development of a series of ideas.  First, in order to find a solution of string theory whose low energy
degrees of freedom are 2+1-dimensional fermions, we shall use a system of D-branes where the number of Neumann-Dirichlet
boundary conditions for the open strings of interest is $\#ND=6$.  In this configuration, supersymmetry is completely broken. The open string spectrum
has no tachyon.  The only massless states are in the Ramond sector,  and are therefore  fermions \cite{Polchinski:1998rq}.
Furthermore, the fermions are chiral in the 2+1-dimensional sense that they have two-components and their mass term
violates the discrete space-time symmetries of parity and time reversal invariance.

This leaves a number of possibilities which are related to each other by T-duality.  The technically simplest
is the $D3/D7$ configuration outlined below
\begin{equation}\label{dbranes}
\begin{array}{rcccccccccccl}
  & & x^0 & x^1 & x^2 & x^3 & x^4 & x^5 & x^6 & x^7 & x^8 & x^9 &\\
& D3 & \times & \times & \times & \times & & &  & & & & \\
& D7 & \times & \times & \times &  & \times  & \times & \times & \times & \times & &   \\
& D5 & \times & \times & \times &  & \times  & \times & \times & &  & &  \\
\end{array}
\end{equation}
The $N_3$ $D3$ and $N_7$ $D7$ branes
are extended in 2+1-spacetime dimensions $(x^0, x^1, x^2)=(t,x,y)$ with
$SO(2,1)$ Lorentz symmetry. The lowest energy states of the 3-7 open strings
are $N_3N_7$ species of 2+1-dimensional 2-component fermions, similar to those in (\ref{fermionlagrangian}).\footnote{The $N$ that we used in eqs.~(\ref{fermionlagrangian})-(\ref{massiveresponse}) would be replaced
by $N_3N_7$.} As a reference system, which will see some use in a later section of this paper, we have also included a $D5$-brane oriented so that it forms a $\#ND=4$ system, {\it i.e.} a BPS state, with the $D3$. This is also the effective $D5$ charge orientation induced on the above $D7$-brane by an internal flux on the $x^7$-$x^8$ plane. Finally, note that the $x^9$ direction is orthogonal to both the $D3$ and $D7$. Thus, the $D3$ and $D7$ can be separated in this direction, introducing a bare mass for 3-7 strings.

The $D3/D7$ system with this configuration was already used by Rey \cite{rey}
as a model of graphene and by other authors to study the quantum Hall plateau transition \cite{Davis:2008nv}.  To apply holography, the number of $D3$-branes $N_3$ is taken as large
and they are replaced by the $AdS_5\times S^5$ geometry. The $D7$-branes are treated as probes and the dynamical problem is to find their embedding in $AdS_5\times S^5$. The embedding used by Rey is
topologically equivalent the product space $AdS_4\times S^4\subset AdS_5\times S^5$.
However, that model  is unstable to  fluctuations of the $D7$ geometry which violate the Breitenholder-Freedman bound for $AdS_4$.
The instability has been interpreted as the existence of a phase transition to a phase with spontaneously broken chiral symmetry which occurs as the coupling constant is increased \cite{rey} (see \cite{kp} and references therein for recent work on this subject).

A way to construct a system which is stable at strong coupling
was suggested by Myers and Wapler \cite{Myers:2008me} who observed that, if one uses
the same embedding geometry, but adds $D7$ world-volume $U(N_7)$ gauge fields in a topological instanton configuration  on $S^4$,
and if the instanton number is large enough, the embedded $D7$-brane becomes stable, at least to small fluctuations of the
geometry. An important further refinement of this idea
was found by Bergman et.~al.~\cite{Bergman:2010gm} who observed that, if instead of $S^4\subset S^5$, one considers $S^5$ as a fibration of two 2-spheres $(S^2,\tilde S^2)$ over an interval $\psi\in[0,\tfrac{\pi}{2}]$, with metric
\begin{align}\label{s5metric}
dS_{S^5}^2 = d\psi^2+\sin^2\psi ds_{S^2}^2+\cos^2\psi d\tilde s_{\tilde S^2}^2
\end{align}
and the world-volume of $D7$ wrapping $S^2$ and $\tilde S^2$, this configuration can be stabilized by adding world-volume gauge
fields with $U(1)$ Dirac monopole fields on one or both of spheres, {\it i.e.} turning on a world-volume $U(1)$ flux with
\begin{equation}
2\pi\alpha' F_0 =\tfrac{R^2 }{2}f \Omega_2
+\tfrac{R^2}{2} \tilde f \tilde\Omega_2
\label{backgroundflux}
\end{equation}
where $(f,\tilde f)$ are constant flux densities and
$(\Omega_2,\tilde\Omega_2)$ are volume 2-forms on the unit spheres $(S^2,\tilde S^2)$ and
where $R$ is the radius of curvature of $AdS_5$.
The Dirac quantization condition is
\begin{equation}  {R^2 \over 2\pi \alpha'} f={n_D\over N_7}~,
 ~{R^2 \over 2\pi \alpha'} {\tilde f}={\tilde n_D\over N_7}\end{equation}
 where $n_D$ and $\tilde n_D$ are integers.
This has the technical simplification of using
Abelian  rather than non-Abelian worldsheet gauge fields.  As well, it allows one to consider the case of a single $D7$-brane,
$N_7=1$,  where the world-volume gauge symmetry is $U(1)$.
The result is a model which allowed the authors of Ref.~\cite{Bergman:2010gm} to produce a beautiful proposal for
holographic duals of quantum Hall states.

A further word is required about the internal fluxes (\ref{backgroundflux}). The $D7$-brane action has the Wess-Zumino coupling
\begin{equation}
\int F \wedge F \wedge C_4
\end{equation}
where $C_4$ is the Ramond-Ramond field in type IIB SUGRA which is sourced by $D3$-branes. Thus a non-zero instanton number density, such as we have in our construction, introduces an effective $D3$ charge on the $D7$-branes. The fluxes (\ref{backgroundflux}) induce an integer $D3$ charge $\Delta N_3 = n_D \tilde n_D$. Since the probe brane carries this charge off to the boundary, this accounts for the shift in the rank of the Yang-Mills gauge group across the defect \cite{Myers:2008me} as depicted in figure \ref{cft}.

In the following, we shall use the construction developed in Ref.~\cite{Bergman:2010gm} with an additional ingredient,
the requirement that the ultraviolet limit of the theory have the discrete symmetries  parity (${\cal P}$) and charge conjugation (${\cal C}$). Both of these are
putatively violated by the construction, ${\cal C}$ by the
background gauge fields $F_0$ and ${\cal P}$ by the Wess-Zumino term in the action for the $D7$-brane. Both are restored by augmenting
them with additional transformations.

${\cal C}$ is easy to fix.  It is defined as $F\to -F$ for all worldsheet gauge field strengths.
It can be augmented by orientation reversing isometries of both of $(S^2,\tilde S^2)$
so that the background field $F_0$ is invariant under the combined transformation.

The dynamical problem of determining the $D7$-brane embedding and its
world-volume gauge fields is obtained from the Dirac-Born-Infeld and Wess-Zumino actions
\begin{align}\label{dbiwz}
S=\frac{ N_7T_7}{g_s}~\int d^8\sigma~\left[\sqrt{\det(g+2\pi\alpha'F)}-\frac{(2\pi\alpha')^2}{2} i F\wedge F\wedge\omega^{(4)}\right]
\end{align}
where $g$ is the world-volume metric, $N_7$ is the number of $D7$-branes
and $T_7=\tfrac{1}{(2\pi)^7{\alpha'}^4 }$ is the D7-brane tension,
$g_s$ is the closed string coupling constant, to which the radius of curvature of $AdS_5$ is
related by  $R^4=4\pi g_s N_3(\alpha')^2$.  The positive  overall sign in (\ref{dbiwz}) and the factor of $i$ in front of the Wess-Zumino
term are due to our use of Euclidean signature.
The
Ramond-Ramond 4-form $\omega^{(4)}$ is defined by $d\omega^{(4)}=\tfrac{4}{R}\omega^{(5)}$ where $\omega^{(5)}$ is
the sum of volume 5-forms of $AdS_5$ and $S^5$.  Generally, a stack of $N_7$ coincident $D$-branes has a $U(N_7)$ gauge
field on its unit volume and $F$ in (\ref{dbiwz}) is the non-Abelian field strength.  Here, we are assuming that only
the $U(1)$ component of this field strength is excited and the rest of the components are set to zero. This assumption
is consistent with the equations of motion for $F$.

We will use the $AdS_5\times S^5$ metric
\begin{align}\label{ads5metric}
  dS^2  =R^2\left[ r^2(dt^2+dx^2+dy^2+dz^2)+\frac{dr^2}{r^2} +dS^2_{S^5}\right]
\end{align}
with $dS^2_{S^5}$ given in (\ref{s5metric}) and
\begin{equation}\label{omega4}
\omega^{(4)}=R^4r^4dt\wedge dx\wedge dy\wedge dz+R^4\frac{c(\psi)}{2}d\Omega_2\wedge d\tilde\Omega_2
\end{equation}
with
\begin{equation}\label{c}
\partial_\psi c(\psi)=8\sin^2\psi\cos^2\psi = 1-\cos 4\psi
\end{equation}

 Parity in 2+1-dimensions is the transformation $(t,x ,y)\to( t,-x ,y)$ which is an orientation reversing isometry
under which the  $AdS_5$-component
of $\omega^{(4)}$ changes sign. Since gauge fields transform covariantly (the background field $F_0$ is invariant), the component of $F\wedge F$ on $AdS_5$ also flips sign.
To make the Wess-Zumino term in (\ref{dbiwz}) parity invariant, we must augment the
orientation reversing isometry of $AdS_5$ with a parity reversing isometry of $S^5$, $\psi\to\tfrac{\pi}{2}-\psi$.
This transformation interchanges $S^2$ and $\tilde S^2$.  It will be a symmetry of the background field
$F_0$ only when the fluxes are equal, $f=\tilde f$.
For fluctuations of the embedding to satisfy the BF bound at large $r$, it is necessary that $f^2\geq\tfrac{23}{50}$ \cite{Bergman:2010gm}.

We take the ansatz for the $D7$-brane which wraps the 2-spheres $S^2$ and $\tilde S^2$,
where $F=F_0$ given in (\ref{backgroundflux}), $f=\tilde f$, it shares coordinates $(t,x,y,r)$ with
$AdS_5$ in (\ref{ads5metric}) and has $r$-dependent positions $z(r)$ and $\psi(r)$, with
induced metric
\begin{align}\label{d7metric}
\frac{ ds_7^2}{R^2} =&r^2(dt^2+dx^2+dy^2)+\frac{dr^2}{r^2}(1+r^4{z'}^2+r^2{\psi'}^2)
+ \cos^2\psi dS_{S^2}^2+\sin^2\psi d\tilde S_{S^2}^2
\end{align}
where $z',\psi'=\tfrac{dz}{dr},\tfrac{d\psi}{dr}$.
The action and equations of motion are
\begin{align}\label{d7action}
S_7
=\tau_7\int dr~\left[r^2\sqrt{(f^2+4\sin^4\psi)(f^2+4\cos^4\psi) }
\sqrt{1+r^4{z'}^2+r^2{\psi'}^2 }
-f^2r^4z'\right]
\end{align}
\begin{align}
\label{soliton}
&\frac{(r\tfrac{d}{dr})^2\psi}{1+(r\tfrac{d}{dr}\psi)^2}+\left(3+
\frac{4 {p_z} {r^{-4}}(f^2+ {p_z}{r^{-4}})}{V}\right)r\tfrac{d}{dr}\psi
= \frac{4\cos2\psi\sin2\psi(\sin^22\psi-f^2)}{V}
\\
\label{z}
&r^2\frac{dz}{dr}=  \frac{(f^2+p_zr^{-4})\sqrt{1+(r\tfrac{d}{dr}\psi)^2}}{\sqrt{V}}\\
&V=(f^2+4\cos^4\psi)(f^2+4\sin^4\psi)-(f^2+\tfrac{p_z}{r^4})^2
\end{align}
with $\tau_7= N_7T_7R^8(2\pi)^2V_{2+1}$ where $V_{2+1}$ is the
volume of the space spanned by the coordinates $(t,x,y)$.  Here, the fact that $z(r)$ is a cyclic variable
in the action (\ref{d7action}) has been used to integrate its equation of motion once and $p_z$ is
the integration constant.

In addition to the problem of the embedding of the $D7$-brane,
we shall need to study the current-current correlation function
in this holographic framework.  The $U(1)$ current is the field theory dual of the $U(1)$ gauge field whose curvature is $F$
in (\ref{dbiwz}).  To get the 2-point function of this current,
 it is sufficient to study fluctuations of the field strength $F=F_0+\tilde F$ about $F_0$ in (\ref{dbiwz}) to quadratic order,
 {\it i.e.} second order in $\tilde F$.
 To that order, it is consistent
to simply insert the solution for the $F=F_0$ geometry into the equation of motion for $\tilde F$. For our purposes, it is
also sufficient (and  consistent with equations of motion)
to take $\tilde F$ to have non-zero components only on the $AdS_5$ space and to only depend on the coordinates
$(r,t,x,y)$.   The action for $\tilde F$ at quadratic order
is
$$
S_M=  N_7T_7(2\pi\alpha')^2 \int d^8\sigma~\left[\frac{1}{4}\sqrt{\det(g+2\pi\alpha' F_0)}~g^{\mu\nu}g^{\lambda\rho}\tilde F_{\mu\lambda}\tilde F_{\nu\rho}-i\frac{1}{2} \tilde F\wedge \tilde F\wedge \omega^{(4)}\right]
$$
In addition to this bulk action, in order to maintain invariance under the RR gauge transformation which would shift
the function $c(\psi)$ defined in equation (\ref{omega4}) by a constant,
it is necessary  to add a surface term to this action \cite{Bergman:2010gm}.  The result
is equivalent to integrating the Wess-Zumino term by parts, and dropping the resulting surface term,
so that the action becomes
\begin{equation}\label{maxwellaction2}
S_M=  N_7T_7(2\pi\alpha')^2 \int d^8\sigma~\left[\frac{1}{4}\sqrt{\det(g+2\pi\alpha' F_0)}~g^{\mu\nu}g^{\lambda\rho}\tilde F_{\mu\lambda}\tilde F_{\nu\rho}-i\frac{1}{2} \tilde A\wedge \tilde F\wedge d\omega^{(4)}\right]
\end{equation}
Then, with our Ans\"atz for
the $D7$ geometry (\ref{d7metric}), and choosing the $A_r=0$ gauge, we obtain
\begin{equation}\label{maxwellaction3}
S_M  =  \frac{N_3N_7}{2\pi^2}\int d^3x\int_0^\infty d\rho \left( \frac{1}{2}(\partial_\rho A_a)^2
+
\frac{1}{4}\alpha^2 F_{ab}^2+i\partial_\rho c(\psi)\epsilon_{rabc}A_a\partial_bA_c\right)
\end{equation}
where we have dropped the tildes on $F$ and $A$, $a,b,c$ are 2+1-dimensional Euclidean indices, $\alpha^2=(f^2+4\sin^4\psi)(f^2+4\cos^4\psi)$
and the radial variable has been transformed to
\begin{equation}\label{rho}
\rho(r)=\int_r^\infty \frac{d\tilde r}{ {\tilde r}^2}\tfrac{\sqrt{(f^2+4\sin^4\psi)(f^2+4\cos^4\psi)}}
{\sqrt{1+\tilde r^2{\psi' }^2+\tilde r^4{z'}^2}}
\end{equation}
Note that $r=0$ and $r=\infty$ are mapped to $\rho=\infty$ and $\rho=0$, respectively.
Using the fourier transform
$$
A_a(x)=\int \frac{d^3q}{(2\pi)^{\frac{3}{2}}}e^{iq\cdot x}A_a(q)
$$
the  field equation can be written
\begin{equation}\label{maxwellequation}
\left[- \partial_\rho^2 +
\alpha^2 q^2
\pm 2iq\partial_\rho c(\psi)\right]A_\pm(\rho,q)=0
\end{equation}
and $q_aA_a =0$. We are considering polarization states which obey
$\left[iq_b\epsilon_{abc}\right]A_{ c\pm}=\pm q A_{a \pm}$.  The field equation should be solved with the
requirement that the solution is regular at the Poincar\'{e} horizon, $\rho=\infty$.
The on-shell action for the gauge field is then
\begin{equation}\label{onshellaction}
\hat S~=~ -\frac{N_7N_3}{4\pi^2}~\int d^3q~{\lim_{\rho\to0}}\left[A_a(\rho,-q)\partial_\rho A_a(\rho,q)\right]
\end{equation}
The current-current correlator is obtained by taking two derivatives of $e^{-\hat S}$ by the
boundary value of the gauge field $A_a(0,q)$ and then setting $A_a(0,q)=0$.  In the following sections we shall quote
the results of this procedure for our three solutions of the worldsheet geometry. Details of our solution
of the Maxwell equation and derivation of current-current correlators are given in the appendix.

\section{Parity and time reversal invariant conformal field theory}

The parity invariant solution of (\ref{soliton}) and (\ref{z}) is
\begin{equation}\psi=\frac{\pi}{4}~~,~~z(r)=z_0-\frac{f^2}{\sqrt{1+2f^2}}\frac{1}{r}
\label{parityinvariantsolution}
\end{equation}
which requires that we set the integration constant $p_z=0$.
The $D7$-brane sits at a constant angle and approaches the Poincar\'{e} horizon at
$r=0$. The world-volume of the $D7$-brane is $AdS_4\times S^2\times S^2$, where the $S^2$'s
have radii $1/\sqrt{2}$ and the radius of
curvature\footnote{These radii are all in units of $R$, the radius of curvature of the ambient $AdS_5\times S^5$.}
of $AdS_4$ is $ {(1+f^2)}/{\sqrt{ 1+2f^2 }}$.
This solution is dual to a conformal field theory which inherits the $SO(3,2)$ symmetry of $AdS_4$.

The flux $f$ remains as a tuneable parameter.   The solution exists independently of the number D7-branes $N_7$,
as long as $N_7<<N_3$.  The dual field theory has a $U(7)$ global symmetry (to apply to graphene, we would set $N_7=4$).

 Note that $z=z_0$ when $r$ is infinite and $z\to-\infty$ as $r\to 0$. We interpret this behavior as the
 brane being a fat interface between two regions in 3+1-dimensional space.  The regions are occupied
 by theories with different densities, being ${\cal N}=4$ Yangs-Mills theories with different rank gauge groups.
 \begin{figure}
 ~~~~~~~~~~~~~~~~~~~~\includegraphics[scale=.6]{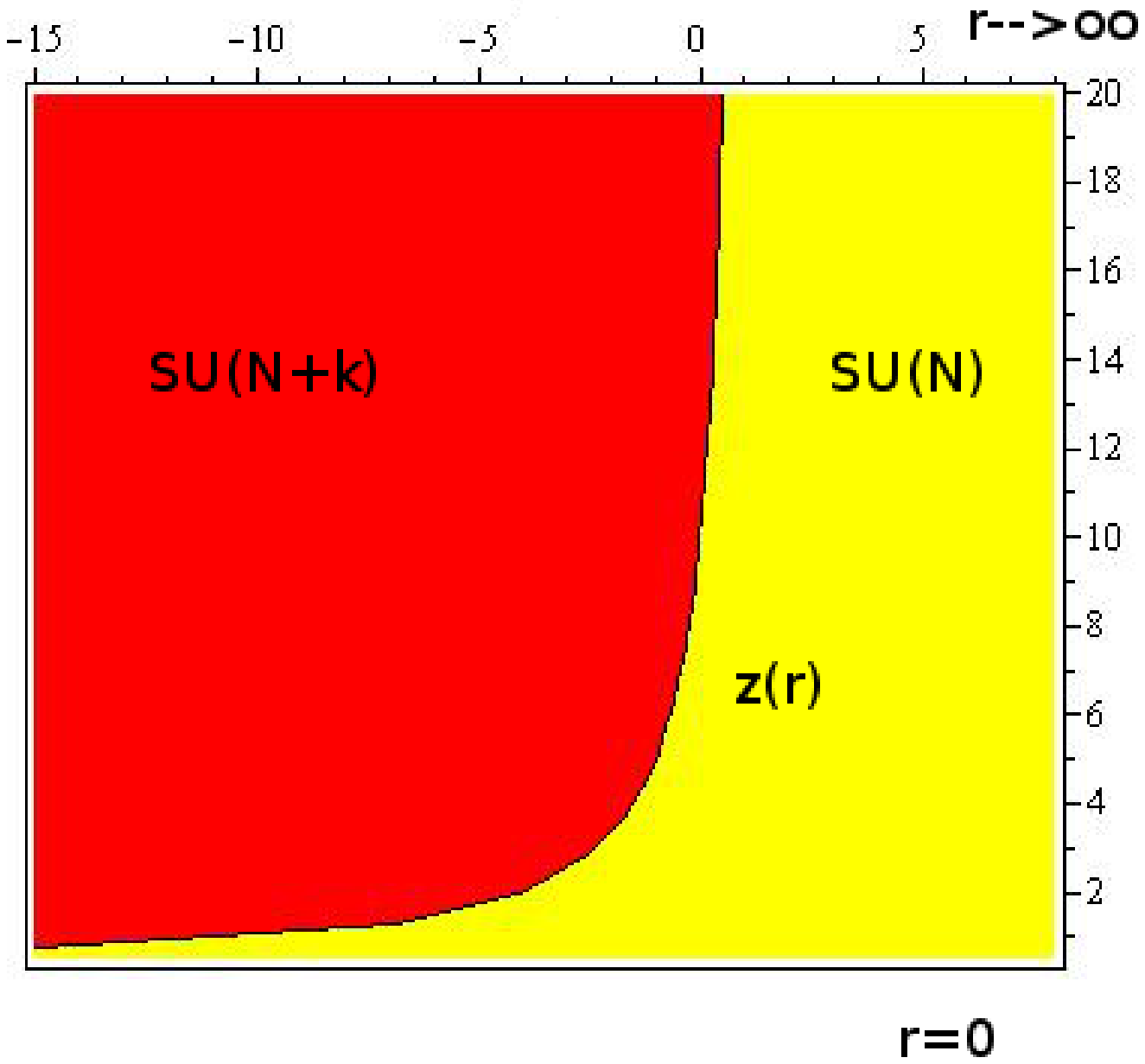}\\
\begin{caption} {The conformal field theory of fermions confined to the 2+1-dimensional woldvolume of a planar defect,
depicted as the vertical curve in the center of the picture, and dividing 3+1-dimensional spacetime into two regions. Here, only the coordinate $z$ which is transverse to the
defect is visible.  It is plotted in the horizontal direction.  The radial AdS coordinate $r$ is plotted in the
  vertical direction.  In the field theory, $r$ corresponds to an energy scale at which one probes the system.
 The region on the left is occupied by ${\cal N}=4$ supersymmetric Yang-Mills theory with $SU(N_3+k)$ gauge group whereas the region on the right is occupied by ${\cal N}=4$ supersymmetric Yang-Mills theory with $SU(N_3)$ gauge group. Here, $k=n_D^2$ is the number of $D3$-branes that are
dissolved into the $D7$-brane. The fact that the interface moves to the left as $r$ decreases from infinity is interpreted as the apparent position of the interface changing as one changes the wave-length at which it is observed, from $z_0$ in the ultraviolet limit ($r\to\infty$), moving all the way to $z=-\infty$ in the infrared limited ($r\to 0$). \label{defectcft} }\end{caption}
\end{figure}
Then, the $r$-dependence of the solution can be interpreted as
 the change in the apparent position of the brane as the energy resolution is varied.
 This is depicted in figure \ref{defectcft}.

Fluctuations about the solution determine the conformal dimensions of operators.  For example,
a solution of (\ref{soliton}) which goes to $\tfrac{\pi}{4}$ at large $r$  must
have the asymptotic behavior
\begin{align}\label{massivelargerasymptotic}
\psi(r\to\infty)=&\tfrac{\pi}{4}+\tfrac{\psi_1}{r^{\Delta_-}}+\tfrac{\psi_2}{r^{\Delta_+}}+\ldots
\end{align}
where
\begin{align}\label{deltaplusminus}
\Delta_{+}=\tfrac{3}{2} + \sqrt{\tfrac{9}{4}-8\tfrac{1-f^2}{2f^2+1}}
~,~\Delta_{-}=\tfrac{3}{2} - \sqrt{\tfrac{9}{4}-8\tfrac{1-f^2}{2f^2+1}}
\end{align}
$\Delta_+$ can be interpreted as the conformal dimension of
a parity violating operator ${\cal O}$, with correlation function fixed by
conformal symmetry,
$$
\left<{\cal O}(x){\cal O}(y)\right>= \frac{\rm const.}{\left|x-y\right|^{2\Delta_+}}
$$
At weak coupling, ${\cal O}$ is identified with $\bar\psi\psi$, the fermion mass operator \cite{Bergman:2010gm}, and
$\left<{\cal O}\right>$ the chiral condensate $\left<\bar\psi\psi\right>$.
However, operator mixing makes this interpretation obscure at strong coupling. The
dimension  $\Delta_{+}$ depends on $f$.  At the bound, $f^2=\tfrac{23}{50}$,
$\Delta_{+}=\tfrac{3}{2}$.
When $f^2=\tfrac{1}{2}$, $\Delta_{+ }=2 $, the classical dimension of $\bar\psi\psi$.  When $f^2=1$
the deformation becomes marginal,  $\Delta_{+ }=3$.  In the interval $f^2\in(\tfrac{9}{14},1)$, $\Delta_->\tfrac{1}{2}$,
the unitarity bound for an operator in 3 dimensions, and it can
also be interpreted as the dimension of an operator in a different conformal field theory with $\psi_1$ being the condensate
and $\psi_2$ the source \cite{Klebanov:1999tb}.

When $\psi=\tfrac{\pi}{4}$, $\alpha$ and $c(\psi)$ are
independent of $\rho$, the Maxwell equation (\ref{maxwellequation}) can be solved exactly and the
current-current correlator extracted by taking two functional derivatives of the on-shell action
(\ref{onshellaction}) by
boundary data $A_a(\rho=0,q)$.
The result is
\begin{equation}
\Delta_{\rm T}= \frac{N_3N_7}{2\pi^2} \frac{f^2+1}{q}~~,~~\Delta_{\rm CS}=0
\label{pinvariantsolution}
\end{equation}
$\Delta_{\rm CS}$ vanishes due to parity symmetry. The function $\Delta_{\rm T}(q)$, whose
dependence on $q$ is determined by conformal symmetry together with the fact that $j_a$ is a conserved current,
 now determines the current-current
correlator at strong coupling.  It is qualitatively similar to the weak coupling result for massless fermions, $\Delta_{\rm T}=\tfrac{N_3N_7}{16\pi q}$ but with a coefficient that depends of $f$.   Recall that $f>\sqrt{\frac{23}{50}}$, and is
typically of order one.

Note that the current-current correlation function is proportional to $N_3$ which is supposed to be taken to infinity.
This limit can make sense if we remember that we have assumed that the quantum of charge in the current is equal to one.
However, in the dual field theory, the Yang-Mills theory has a coupling constant $g_{\rm YM}$ and it would be more sensible
to normalize the charge by taking into account this coupling constant.  Then the current-current correlation functions
would have the factor $N_3$ replaced by $N_3g_{\rm YM}^2\equiv \lambda$,
the 't Hooft coupling,
which is held fixed in the large $N_3$ limit. It is natural that the current-current correlator is proportional to $\lambda$
at weak coupling.  The holographic description that we have been discussing here is accurate when $\lambda$ is
large and our
result indicates that the correlation function still is proportional to $\lambda$ in the strong coupling limit.

The conductivity in this limit is given by (see reference \cite{Myers:2008me} for a detailed discussion)
$$
\sigma_{xx}=  \frac{N_7\lambda(f^2+1)}{2\pi^2}   \sqrt{1-\vec q^2/\omega^2}
$$
where we have gone back to Lorentzian signature space-time- and $q_\mu=(\omega,\vec q)$.
When the wave-vector is put to zero, $\sigma_{xx}$ approaches a constant.

\section{Parity and time reversal violating solution}

At small $r$, a solution  of (\ref{soliton}) must have the form
\begin{align}\psi =\tfrac{\arcsin f}{2}+cr^{\nu}+\ldots~,~
\label{massivesmallrasymptotic}
\nu=
\sqrt{\tfrac{9}{4}+16\tfrac{1-f^2}{4-f^2}}
-\tfrac{3}{2}
\end{align}

\begin{figure}
 ~~~~~~~~~~~~~~~~~~~~\includegraphics[scale=.7]{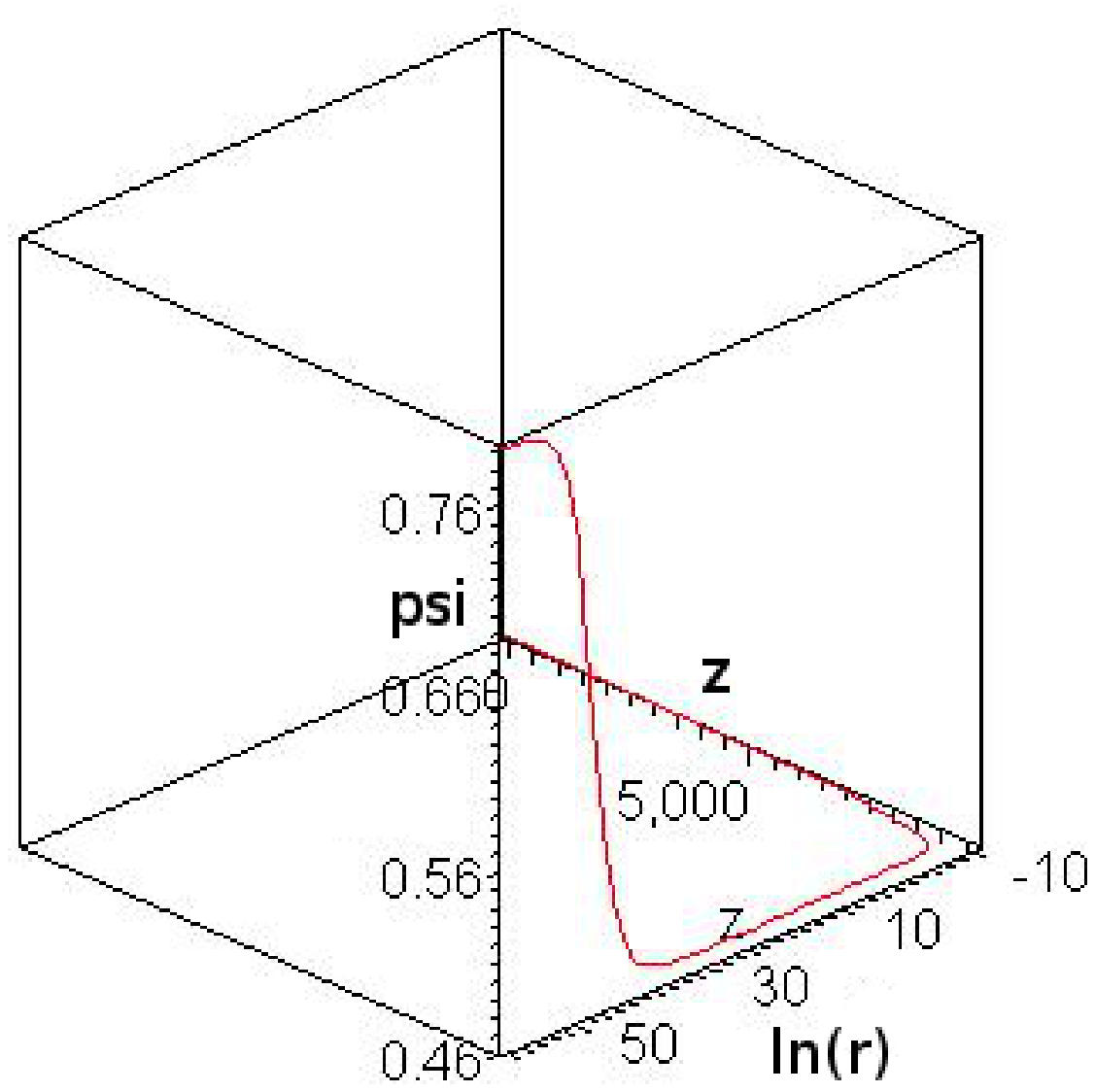}\\
\begin{caption} {Numerical solution of (\ref{soliton}) and (\ref{z}) with $f=.8$.
  $\psi$ is plotted on the vertical axis, $\psi( \infty)=\tfrac{\pi}{4}$ and $\psi(0)=\tfrac{1}{2}\sin^{-1} f\approx 0.46$.  $z$ is plotted on the rear axis and we see that $z\sim 1/r$ for small $r$ and $z\sim$constant at large $r$.\label{running}}\end{caption}
\end{figure}

If we search for a solution where $\psi$ depends on $r$ with boundary behavior (\ref{massivelargerasymptotic}) and (\ref{massivesmallrasymptotic}), once one
of the three constants $(\psi_1,\psi_2,c)$  is fixed, the other two are determined by requiring  that the solution
is nonsingular.
(Solutions at $r\sim0$ exist only when $p_z=0$.)
Setting $\psi_1$ to some fixed value corresponds to turning on a source for the operator ${\cal O}$ with conformal dimension $\Delta_+$,
and whose expectation value $\left<{\cal O}\right>$ is then
proportional to the other constant  $\psi_2$.  Since $\Delta_+$ is positive, the operator is relevant and the
dual field theory is no longer a conformal field theory.   However, it will flow to another conformal field theory
in the infrared, small momentum limit.  The
equations (\ref{soliton}) and (\ref{z})
can be solved numerically.
An example of a solution is depicted in figure \ref{running}.

For the solution in figure \ref{running}, where $\psi$ is not constant, we cannot obtain an exact solution of the Maxwell equation (\ref{maxwellequation}).  However, it is
easy to solve in the limits where $q$ is large or small compared to other dimensional parameters.  Here, the only dimensional
parameters in the problem come from the boundary behavior and we can use,
for example, $\psi_1^{1/\Delta_-}$ to compare with $q$.

When $q$ is
large, as is shown in the Appendix,
the solution is identical to (\ref{pinvariantsolution}).  The theory has an ultraviolet fixed point which
is identical to the parity and time reversal invariant conformal field theory that corresponds to the constant $\psi=\tfrac{\pi}{4}$
solution.

On the other hand, when $q$ is small, we find
\begin{equation}
\Delta_{\rm T}= \frac{N_3N_7}{2\pi^2}\frac{2f}{q}+\ldots~~,~~\Delta_{\rm CS}=  \frac{N_3N_7}{\pi^2} (f\sqrt{1-f^2}-\arccos f)
+\ldots
\label{pnoninvariantsolution}
\end{equation}
where corrections are of higher order in $q$.
These functions characterize the electromagnetic properties of the infrared fixed point of the field theory dual
of the non-constant  $\psi$ solution.  That theory is apparently gapless and parity and time reversal violating.
We note that $\Delta_{\rm CS}(0)$ differs from the one-loop result (\ref{massiveresponse}).

By doubling the degrees of freedom, we could find a parity invariant configuration. This could be, for example,
an exotic time reversal
invariant phase of graphene.  In that case, two D7 branes behave as we have described above and two D7-branes
are their parity mirror, with $\psi(r)$ replaced by $\tfrac{\pi}{2}-\psi(r)$.  Since the pairs of branes behave differently,
the global $U(4)$ symmetry is broken to $U(2)\times U(2)$.
In this case, this is not spontaneous breaking, instead it is explicit
breaking by turning on the boundary condition (\ref{massivelargerasymptotic}) for two of the D7-branes and a similar
boundary condition but with $\psi_1,\psi_2$ replaced by $-\psi_1,-\psi_2$ for the other two D7-branes. At weak coupling,
this would correspond to turning on the parity invariant mass that corresponds to a charge density wave in graphene
which was discussed in reference \cite{Semenoff:1984dq}. At weak coupling, this would seem to gap the spectrum, but for
this particular solution, at strong coupling gapless charged excitations seem to survive.
The parity even part of the current-current correlator $\Delta_{\rm T}$ would be as we computed above
with $N_7=4$, however $\Delta_{\rm CS}$ would cancel, and would therefore be zero, consistent with parity and time reversal invariance.  On the other hand, it would re-appear in a flavor current correlation function, and it would give the so-called
``valley Hall effect'', originally described in reference \cite{Semenoff:1984dq}.
The value of $\Delta_{\rm CS}$ in (\ref{pnoninvariantsolution}) with $N_7$ set to 4 describes the valley Hall effect at strong coupling.

\section{Solution with a charge gap}

It is interesting to ask whether we can find a theory with a charge gap.  Charged particles are the low energy modes
 of strings which are suspended between the Poincar\'{e} horizon and the $D$7-brane and the mass of a string state is roughly proportional to its length.   In the previous two examples, the charged
 particles were massless, corresponding to the fact that the $D7$-brane comes arbitrarily close to the Poincar\'{e} horizon
 and the 3-7 strings could be arbitrarily short.  To get a gapped solution, we need a configuration of $D7$-brane which does
 not approach the Poincar\'{e} horizon, something similar to
a ``Minkowski embedding'' where the $D7$ brane pinches off before it reaches the Poincar\'e horizon at $r=0$.

The $D7$-brane can pinch off smoothly when one of the $S^2$'s collapses, for example, when $\psi\to0$.  However, this would not be compatible with the existence of the magnetic flux $f$ on $S^2$ unless there is a magnetic source.
Here, a magnetic source would be supplied by $n_D$ $D5$-branes where $n_D$ is the number of units of monopole flux.  The flat space configuration of the $D5$-branes is given in the table  in (\ref{dbranes}). On $AdS_5\times S^5$,
each $D5$ brane wraps $\tilde S^2\subset S^5$ and has $n_D$ units of Dirac monopole charge on $\tilde S^2$.

However, it can be argued that such a $D7$-brane with a shrinking $S^2$ attaching to a suspended $D5$-brane is not stable, the
$D7$ always has smaller tension than the $D5$.  If the $D5$ where connected to the horizon, it would simply pull the $D7$ brane to
the horizon.  This would seem to rule out the possibility of finding gapped solutions using $D5$-branes suspended between
the $D7$-brane and the Poincar\'{e} horizon.

\begin{figure}
 ~\includegraphics[scale=.7]{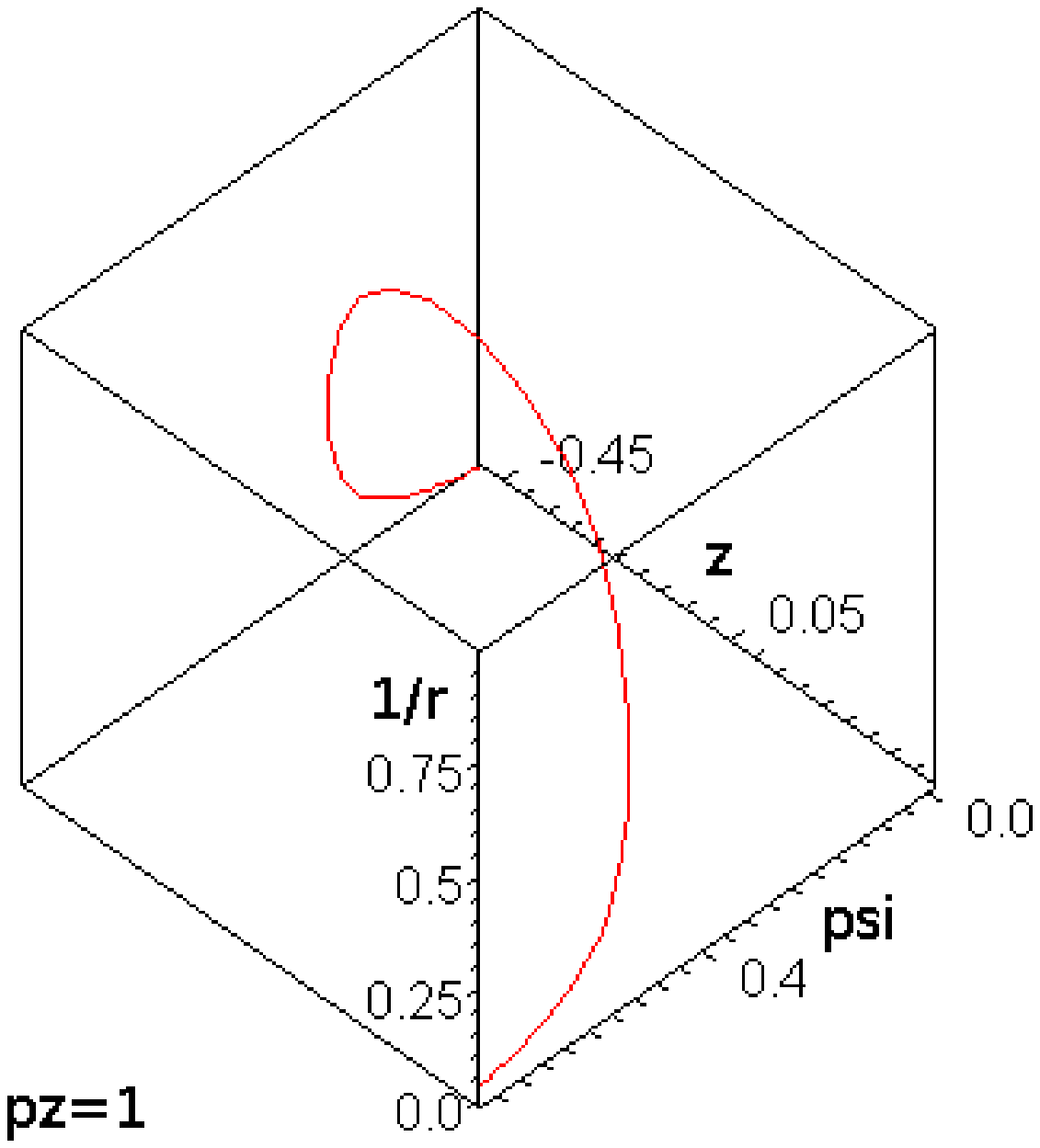}\\
\begin{caption} { Numerical solution of  (\ref{soliton}) and (\ref{z}) with $f=.8$ and $p_z=1$.
  $\psi$ is plotted on the vertical axis. It goes smoothly between $(1/r,\psi,z)=(0,\tfrac{\pi}{4},0.46)$ and $(1/r,\psi,z)=(0,0,-0.46)$.   \label{suspended}
}\end{caption}
\end{figure}

The only alternative is that the $D5$ brane is connected to $r\to\infty$
and that the imbalance of tensions pulls the $D7$-brane back to infinity.  We can indeed find numerical solutions of the $D7$ equation which behave in this way. An example
is depicted in figure \ref{suspended}. The
$D7$-brane begins at $(r,\psi,z)=(\infty,\tfrac{\pi}{4},-0.46)$.  As $\psi$ decreases, $r$ decreases until it reaches a minimum.
Then it starts increasing and returns to the asymptotic region at $(r,\psi,z)=(\infty,0,0.46)$.
During this interval, the coordinate $z$ increases steadily over a  finite range.   At the latter endpoint, $S^2$ has
 collapsed to a point, leaving a source with $n_D$ Dirac monopoles on the $D7$ world-volume.   We can think of the flux on the collapsing $S^2$ as being sourced by
 a $D5$-brane sitting at $r=\infty$.

The solution depicted in figure \ref{suspended}
is fundamentally different from the solutions that we have found in the previous  two Sections
since it goes to the boundary at two different
locations. On the $AdS_5$, these locations are separated in the coordinate $z$, so they nominally correspond to two different
dual quantum field theories.

 The operator
corresponding to fluctuations of $\psi$, for example, in the asymptotic regime where $\psi\to\tfrac{\pi}{4}$
 behaves   as in equation (\ref{massivesmallrasymptotic}), whereas in the regime where $\psi\to0$ it
has $\Delta_-=1,\Delta_+=2$, independent of $f$ and identical to the quantities for a $D5$-brane with flat space
configuration depicted in (\ref{dbranes}).

In addition, to solve the Maxwell equation on the world-volume requires
two sets of asymptotic data.  For example, if we require that the world-volume gauge field goes to
$A_a(x)$ at the $\psi\to\tfrac{\pi}{4}$ asymptote and $\tilde A_a(x)$ at the $\psi\to 0$ asymptote,
it is straightforward to solve the
 large momentum limit where we obtain two decoupled currents (see the Appendix for the details)
 \begin{align}
 \left<j_aj_b\right>&= \frac{N_3N_7}{2\pi^2}\frac{f^2+1}{q}\left(q^2\delta_{ab}-q_aq_b\right)
 ~,~ \left<j_a\tilde j_b\right>=0 \nonumber \\
 \left<\tilde j_a\tilde j_b\right>&= \frac{N_3N_7}{2\pi^2}\frac{f\sqrt{f^2+4}}{q}\left(q^2\delta_{ab}-q_aq_b\right)
 \label{gappedcorrelatorsathighenergy}\end{align}
 The $j-j$ correlator reproduces the high energy limit (\ref{pinvariantsolution}) of our previous solutions. The
 $\tilde j-\tilde j$ correlator produces what would be expected for the solution of the $D5$-brane geometry with
 the constant angle $\psi=0$. The field theory dual to this the $D5$-brane geometry is well known \cite{Myers:2008me}.

 On the other hand, a small momentum expansion is diagonalized by linear combinations of the currents with
 \begin{align}
 &\left<j_{+a}j_{+b}\right>= \frac{N_3N_7}{4\pi}\epsilon_{acb}q_c+\ldots
 \label{jplus}\\
 &\left<j_{-a}j_{-b}\right>=  \frac{N_3N_7}{ \pi^2\rho_m}( \delta_{ab}-\tfrac{ q_aq_b}{  q^2})
 + \epsilon_{acb}q_c \Delta^{(-)}_{\rm CS}(0)+\ldots
\label{jminus}
\end{align}
where \begin{align}
&\rho_m=\int_{r_{\rm min}}^\infty \frac{d\tilde r}{ {\tilde r}^2}\tfrac{\sqrt{(f^2+4\sin^4\psi)(f^2+4\cos^4\psi)}}
{\sqrt{1+\tilde r^2{\psi' }^2+\tilde r^4{z'}^2}} \label{rmin}\\
&\Delta^{(-)}_{\rm CS}(0)=\frac{N_3N_7}{\pi^2} \int_0^{\pi/4} d\psi  (1-\cos 4\psi ) \left(1-\frac{\rho(\psi)}{\rho_m} \right)^2
\end{align}
where  $r_{\rm min}$ in (\ref{rmin}) is
 the minimum value of $r$ that the solution reaches.
 Corrections in (\ref{jplus}) and (\ref{jminus}) are of order $q^2$.

 \begin{figure}
 ~~~~~~~~~~~~~~~~~~~~\includegraphics[scale=.6]{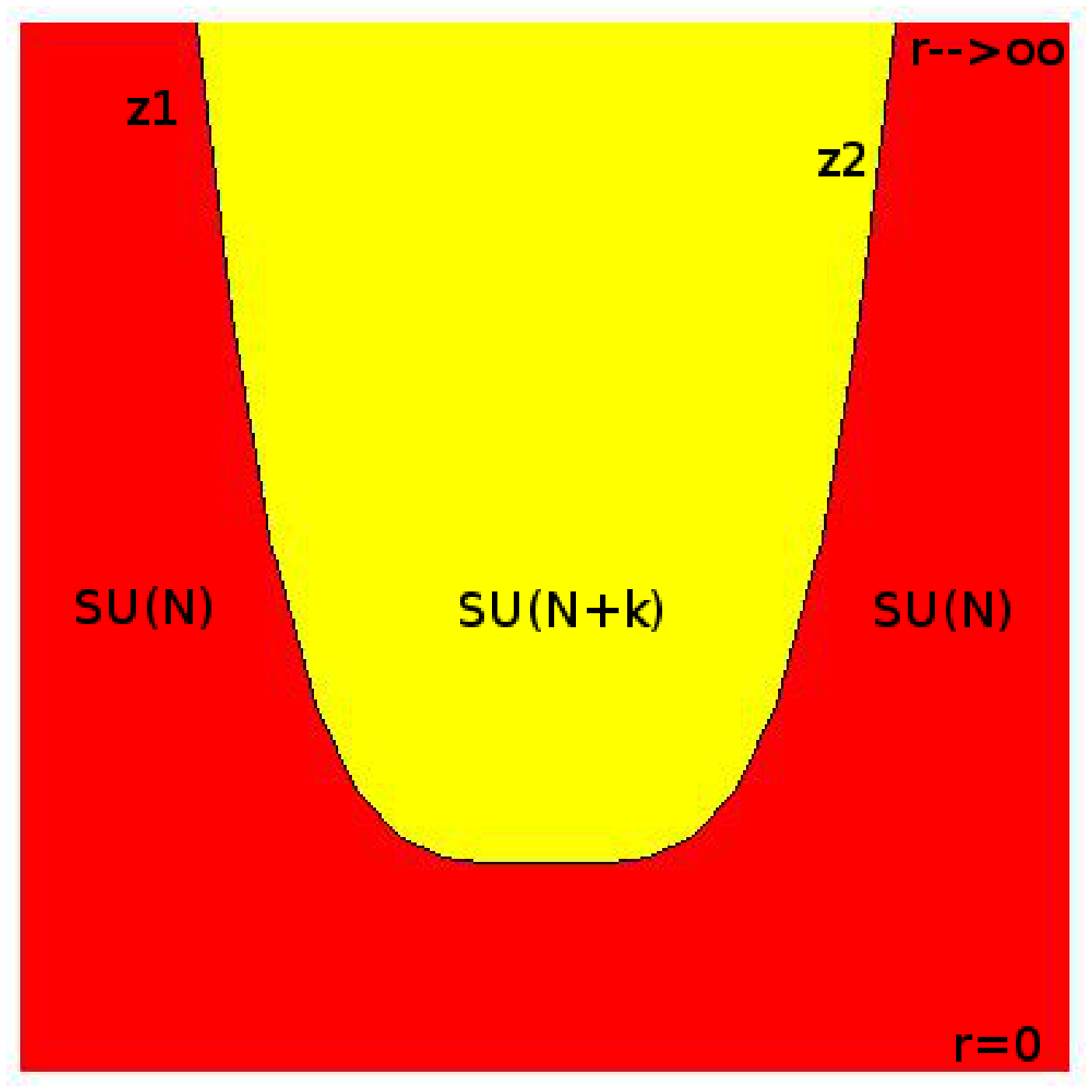}\\
\begin{caption} {The field theory which is dual to the gapped solution has a sandwich containing
$SU(N+k)$ supersymmetric Yang-Mills theory in the interior, sandwiched between two planar defects, one corresponding to the D7-branes and the other corresponding to the D5-branes, each with their own set of degrees of freedom,
 and with $SU(N)$ Yang-Mills theory on the exterior. The field theories on the planar defects are highly
 correlated.
Here, $r$ is the vertical and $z$ the horizontal directions. The width of the sandwich is the range of
$z$ which is finite in the ultraviolet, large $r$ regime, but shrinks and disappears at a scale which is
the minimum of $r$.   \label{hanging}}\end{caption}
\end{figure}

 The currents $j_{\pm}$ are linear combinations of $j,\tilde j$ which are normalized so that the charges obtained
 by integrating the time components of $j_\pm$ are integers, as were the charges of $j,\tilde j$.  We then observe
 that the $j_+$ current has the value of $\Delta_{\rm CS}(0)$ that would be expected for a
 system of $N_3N_7$ fermions with a parity violating mass gap. This is identical to the non-interacting one-loop
 result for massive fermions.

 Note that $j_-$ has a superfluid-like pole in its 2-point function, indicating
 spontaneous breaking of a phase symmetry.  This breaking is what would be expected from joining of the $D7$ and $D5$-branes.
Figure \ref{hanging} depicts the geometry of the field theory configuration.

\section{Remarks}

A few remarks about the solutions that we have found are in order.  First, if we begin with the
$D7$-brane embedding equation (\ref{soliton}), which we recopy here for the reader's convenience,
$$
\frac{(r\tfrac{d}{dr})^2\psi}{1+(r\tfrac{d}{dr}\psi)^2}+\left(3+
\frac{4 {p_z} {r^{-4}}(f^2+ {p_z}{r^{-4}})}{V}\right)r\tfrac{d}{dr}\psi
= \frac{4\cos2\psi\sin2\psi(\sin^22\psi-f^2)}{V}
$$
$$
V=(f^2+4\cos^4\psi)(f^2+4\sin^4\psi)-(f^2+\tfrac{p_z}{r^4})^2
$$
We recall that, to find a solution that extends to the Poincar\'{e} horizon at $r=0$, we must set
$p_z=0$. We can re-write this equation as
\begin{equation}
r\frac{d}{dr}\ln\left[\frac{(f^2+4\cos^4\psi)(f^2+4\sin^4\psi)-f^4}{ 1+\left(r\frac{d}{dr}\psi\right)^2}\right]
=6\left( r\frac{d}{dr}\psi\right)^2
\end{equation}
The right-hand-side is positive and we can conclude that the left-hand-side is a monotonically
increasing function of $r$.

If we assume that the logarithmic derivative of $\psi$ vanishes at both limits, $r\to\infty$ and $r\to 0$,
integrating this equation yields the sum rule
\begin{equation}
 \frac{(f^2+4\cos^4\psi(\infty))(f^2+4\sin^4\psi(\infty))-f^4}{(f^2+4\cos^4\psi(0))(f^2+4\sin^4\psi(0))-f^4}
=\exp\left( 6\int _0^\infty \frac{dr}{r}\left( r\frac{d}{dr}\psi\right)^2\right)
\label{sumrule}
\end{equation}
This indicates that any solution of the equation of motion will necessarily have smaller $V=(f^2+4\sin^4\psi)(f^2+4\cos^4\psi)
 - f^2$ at $r=0$, that is in
the infrared limit, than at $r=\infty$, the ultraviolet limit.  If we interpret the evolution from large $r$ to
small $r$ as a renormalization group flow, we have found a quantity which definitely decreases.  What is more, it
is directly related to the Routhian\footnote{This is obtained by the Legendre transform $R(\psi(r), \psi'(r), p_z) = L(\psi(r),\psi'(r), z'(r)) - z'(r) p_z$ and setting $p_z=0$ to focus on embeddings entering the Poincar\'{e} horizon.}
\begin{align}
 {\cal R}_7 =\frac{N_3N_7V_{2+1}}{2\pi^2}\int_0^\infty dr~r^2 \sqrt{(f^2+4\sin^4\psi)(f^2+4\cos^4\psi)
 - f^2 }
\sqrt{1+\left(r\frac{d}{dr}\psi\right)^2}
\end{align}
whose variation gives the equation of motion.  It is the density in this integral which must decrease, an analog
of an H-theorem which has recently been discussed for nonsupersymmetric 2+1-dimensional field theories \cite{Klebanov:2011gs}.

If we evaluate the left-hand-side of (\ref{sumrule}) with the solution that we have found, it reads
\begin{equation}
 \frac{1+2f^2}{f^2(4-f^2)}
=\exp\left( 6\int _0^\infty \frac{dr}{r}\left( r\frac{d}{dr}\psi\right)^2\right)
\label{sumrule}
\end{equation}
The left-hand-side is greater than one for all values of $f^2<1$ and it equals one when $f^2=1$.
When $f\to 1$, the operator that we have perturbed the conformal field theory by when we switched on
$\psi_1$ approaches a marginal operator.  It should be possible to develop a perturbative approach
where this flow can be analyzed explicitly.

As another observation, in the same solution, the scale symmetry of the embedding equation for the
$D7$-brane is broken only by one of the parameters in the boundary condition, say $\psi_1$.  Then the
other parameters in the boundary conditions are determined by $\psi_1$ and the relationship is
governed by scale symmetry,
$$
\psi_2= g(f)\psi_1^{\Delta_+/\Delta_-}
$$
It would be interesting to compute the function $g(f)$ to see if it has any special behavior.

Finally, the gapped solution that we have found in section 5 seems to be the unique example of a simple solution
with a mass gap.
We have outlined why it is unlikely that a solution could exist with the $D5$-brane reaching the Poincar\'{e} horizon, rather
than being located at the large $r$ boundary.
We believe that the argument is robust.
We cannot rule out more complicated complexes of branes with multiple intersections.
The solution that we did find had an additional asymptotic region which essentially doubles the degrees of
freedom.   It also has a spontaneously broken $U(1)$ symmetry which we find explicitly when we study the current-current
correlation functions.   This breaking could proceed with a $\bar\psi\tilde\psi$-condensate which breaks the $U(1)\times U(1)$
gauge symmetry to a diagonal $U(1)$. This would be sufficient to gap the spectrum of fermions which live on both the $D7$ and $D5$ branes.

\acknowledgments
 This work is supported by NSERC of Canada and in part by the National Science Foundation
 under Grant No.~NSFPHY05-51164. G.W.S.~acknowledges the hospitality of the KITP in Santa Barbara
 as well as
the Aspen Center for Physics,   Galileo Galilei Institute
and Nordita, where parts of this work were completed.

\appendix

\section{Solution of the world-volume gauge field}

The equation of motion for the worldsheet gauge field $A_{\pm}$ is given in equation (\ref{maxwellequation}) as
$$
\left[-\partial_\rho^2 +q^2\alpha^2(\rho)\pm  iq\partial_{\tilde\rho}(2 c(\rho))\right]A_{\pm}=0
$$
In this Appendix we will outline our solution of this equation for the three different worldsheet geometries
that we have discussed, beginning with the simplest one with constant angle $\psi$.

\paragraph{Constant $\alpha$ and $c$:}

In this case, we have
\begin{align}
\left[-\partial_{  \rho}^2 + q^2\alpha^2  \right]A_{a}=0
\nonumber\end{align}
where $\alpha=f^2+1$. The  general solution is
$$
A= c_1 e^{q\alpha\rho}+c_2 e^{-q\alpha\rho}
$$
The solution must converge at $\rho\to\infty$ (this is $r\to 0$, the Poincar\'{e} horizon).  To
satisfy this condition, we set $c_1=0$.
Then the solution is
$$
A_a(\rho,q)= A_a(q)e^{-q\alpha\rho}
$$
Now, we must plug this solution into the on-shell action (\ref{onshellaction}) to get
$$
\hat S=  \frac{N_3N_7}{4\pi^2} \int d^3q A_a(-q)~\frac{\alpha}{q}(q^2\delta_{ab}-q_aq_b)A_b(q)
$$
Taking two functional derivatives by $A_a(q)$ yields
$$
\Delta_{\rm CS}=0
$$
$$
\Delta_{\rm T}=\frac{N_3N_7}{2\pi^2}\frac{f^2+1}{q}
$$
which is the result quoted in equation (\ref{pinvariantsolution}).

\paragraph{Non-constant $\psi$ and large $|q|$:}

Consider the Maxwell equation at large $q$ where now $\alpha$ depends on $\rho$,
$$
\left[-\partial_\rho^2 +q^2\alpha^2(\rho)\right]A_{\pm}=0
$$
and consider the WKB Ansatz
$$
A=A_0(q)e^W
$$
so that the differential equation becomes
$$
 -W''-(W')^2 +q^2\alpha^2(\rho)  =0
$$
For large $q$, $W\sim q$ and
$$
W=-\int_{ 0}^\rho d\tilde \rho q \alpha(\tilde \rho)
$$
We have chosen the minus sign so that the solution converges at $\rho\to\infty$.
 The on-shell action is
 $$
 S= \frac{N_3N_7}{4\pi^2}\int d^3q W' A_0(q)A_0(-q) = \frac{N_3N_7}{4\pi^2}\int d^3q A_a(-q)\frac{(f^2+1)}{|q|}(q^2\delta_{ab}-q_aq_b)A_b(q)
 $$
 This gives $\Delta_{\rm T},\Delta_{\rm CS}$ identical to the constant solution that we
 analyzed in the above subsection (as we expected).

\paragraph{Non-constant $\psi$ and small $q$:}

We can re-scale $\rho$ in the Maxwell equation by a factor of $q$: $\rho\to \rho/q$ to obtain
$$
\left[-\partial_\rho^2 + \alpha^2(\rho/q)+ i \partial_{ \rho}(2 c(\rho/q))\right]A_{+}=0
$$
and a similar equation for $A_-$ which is the complex conjugation.

From the numerical solution in figure \ref{running}, we see that the interpolation
of $\psi(r)$ between its two asymptotic regions occurs in a relatively small interval. In the small $q$ limit,
this interval is squeezed into the small $ \rho $ region.  To deal with this, we approximate $\alpha$
and $c$ by step functions at a given value of $ \rho_0$.  Then we will take the limit as $ \rho_0$ goes
to zero.

In the  $ \rho< \rho_0$ regime, $\alpha=(f^2+1)^2$.  In the  $ \rho
> \rho_0$ region,
$\alpha^2=4f^2$ and we approximate $\alpha^2$ by the step function
$$\alpha^2 = (f^2+1)^2\theta(\rho_0-\rho)+(2f)^2\theta(\rho-\rho_0)$$
We will approximate the term with a derivative of $c(\psi)$ by a delta function.  Since
$$c(\infty)-c(0)=\int^{\frac{1}{2}\arcsin f}_{\frac{\pi}{4}}
d\psi \partial_\psi c
= \frac{1}{2}\arccos f -\frac{1}{2}f\sqrt{1-f^2}$$
we write
$$
\partial_{ \rho}(2c(\psi))= \left[ \arccos f -f\sqrt{1-f^2}\right]\delta(\rho-\rho_0)
$$
The solution of the equation is then
$$
A_+=\left[ A_1e^{(f^2+1)\rho}+A_2e^{-(f^2+1)\rho}\right]\theta(\rho_0-\rho)
+  B e^{-2f\rho} \theta(\rho-\rho_0)
$$
where we have imposed convergence at large $\rho$.
Continuity of the wavefunction at $\rho=\rho_0$ implies
$$
  A_1e^{(f^2+1)\rho_0}+A_2e^{-(f^2+1)\rho_0} =
 B e^{-2f\rho_0}
$$
and the discontinuity of the derivative due to the delta-function potential requires
$$
  (f^2+1)\left[ A_1e^{(f^2+1)\rho_0}-A_2e^{-(f^2+1)\rho_0}\right]
 -2f B e^{-2f\rho_0}
$$
$$
=- i\left[ \arccos f -f\sqrt{1-f^2}\right]\left[ A_1e^{(f^2+1)\rho_0}+A_2e^{-(f^2+1)\rho_0}\right]
$$
From these equations, we can obtain

$$
A_1-A_2= \left[-\frac{2f}{f^2+1}-i\frac{f\sqrt{1-f^2}-\arccos f}{f^2+1}\right](A_1+A_2)
$$

Then, the on-shell action is
$$
\hat S=-\frac{N_3N_7}{4\pi^2} \lim_{\rho\to 0} (A\partial_\rho A)=-\frac{N_3N_7}{2\pi^2}\frac{f^2+1}{2}(A_1+A_2)(A_1-A_2)$$
$$
=\left[ f+i (f\sqrt{1-f^2}-\arccos f) \right](A_1+A_2)^2
$$

The current-current correlator components are
$$
\Delta_{\rm T}=  \frac{N_3N_7}{2\pi^2}\frac{2f}{q}
$$
$$
\Delta_{\rm CS}= \frac{N_3N_7}{2\pi^2}(f\sqrt{1-f^2}-\arccos f)
$$
as quoted in eq.~(\ref{pnoninvariantsolution}).

 \paragraph{Solution with a charge gap at large $q$:}

  We use a variable $s=\rho$ for the interval $\rho\in[0,\rho_{m}]$ and $s=2\rho_{ m}-\rho$ for $\rho\in
 [\rho_{ m},0]$. The differential equation is
 $$
\left[-\partial_s^2 +q^2\alpha^2(s)\pm  iq\partial_{s}(2 c(\psi(s)))\right]A_{\pm}=0
$$
There are two boundaries, one at $s=0$, the other at $s=2\rho_m$.
Let us assume that $A(0,q)=A(q)$ and $A(2\rho_m,q)=\tilde A(q)$.

We first study the large $q^2$ regime.  The
solution is
$$
A(s)= \frac{-Ae^{-2\hat W}+\tilde A e^{-\hat W}}{1-e^{-2\hat W}}~e^W+\frac{A-\tilde A e^{-\hat W}}{1-e^{-2\hat W}}~e^{-W}
$$
where $W[s]=\int_0^sds'\alpha(s')q$ and $\hat W=W[2\rho_m]$.
Since $W\sim q$ we should take the limit where $W$ is large.
$$
\partial_sA(0)=- (f^2+1)\coth\hat W A\sim -(f^2+1)A
$$
$$
\partial_sA(2\rho_m)=f\sqrt{f^2+4}\coth\hat W \tilde A\sim  f\sqrt{f^2+4}  \tilde A
$$
Then the on-shell action is
$$
\hat S =  \frac{N_3N_7}{2\pi^2}\left[\frac{f\sqrt{f^2+4} }{2q}\tilde A_a (q^2\delta_{ab}-q_aq_b) \tilde A_b
+\frac{(f^2+1)  }{2q}A_a (q^2\delta_{ab}-q_aq_b) A_b\right]
$$
which reveals the result quoted in eq.~(\ref{gappedcorrelatorsathighenergy}).

\paragraph{Solution with a charge gap at  small $q$:}

First, we observe that, when $q\to 0$ the Maxwell equation is solved by
$$
A(s)=A(q) (2\rho_m-s)/2\rho_{\rm max} + s/2\rho_{m}\tilde A(q)
$$
and the on-shell action is proportional to
$$
\left. \frac{1}{2}A\partial_s A\right|^{s=2\rho_m}_0 = \frac{1}{4\rho_m} |\tilde A(q)-A(q) |^2
$$
This is a Higgs-like term, it implies that the difference of currents $j-\tilde j$ gets a pole in its
two point correlator.

   Now, to expand the solution in $q$, it is convenient to convert the differential equation to an integral
   equation,
 $$
 A=A(q) \left( 1-\frac{s}{2\rho_{\rm m}}\right) + \frac{s}{2\rho_{m}}\tilde A(q) +\int_0^s ds'\int_0^{s'}ds'' \left[q^2\alpha^2(s'')\pm  iq\partial_{s''}(2 c(s''))\right]A(s'')
 $$
 $$
  -\frac{s}{2\rho_m}\int_0^{2\rho_m}ds' \int_0^{s'}ds'' \left[q^2\alpha^2(s'')\pm  iq\partial_{s''}(2 c(s''))\right]A(s'')
 $$
 and then we can solve iteratively.  We shall only examine the solution to leading order in $q$.
 Iteration yields
 $$
 A(s)=A  \left( 1-\frac{s}{2\rho_{\rm m}}\right) + \frac{s}{2\rho_{m}}\tilde A  \pm  iq\int_0^s ds'\int_0^{s'}ds'' \partial_{s''}(2 c(s''))  \left[A  \left( 1-\frac{s''}{2\rho_{\rm m}}\right) + \frac{s''}{2\rho_{m}}\tilde A\right]
 $$
 $$
  \mp  iq\frac{s}{2\rho_m}\int_0^{2\rho_m}ds' \int_0^{s'}ds'' \partial_{s''}(2 c(s'')) \left[A  \left( 1-\frac{s''}{2\rho_{\rm m}}\right) + \frac{s''}{2\rho_{m}}\tilde A\right]+\ldots
 $$
 Then
 $$
 \partial_sA(0)=  \frac{1}{2\rho_{m}}(\tilde A-A)
  \mp  iq\int_0^{2\rho_m}ds'(1-\frac{s'}{2\rho_m}) \partial_{s'}(2 c(s')) \left[A  \left( 1-\frac{s'}{2\rho_{\rm m}}\right) + \frac{s'}{2\rho_{m}}\tilde A\right]+\ldots
 $$
 $$
 \partial_sA(2\rho_m)= \frac{1}{2\rho_{m}}(\tilde A-A)  \pm  iq \int_0^{2\rho_m} ds' \frac{s'}{2\rho_{m}}\partial_{s'}(2 c(s'))  \left[A  \left( 1-\frac{s'}{2\rho_{\rm m}}\right) + \frac{s'}{2\rho_{m}}\tilde A\right]+\ldots
 $$

 The on-shell action is proportional to
 $$
 \tilde A\partial_sA(2\rho_m)-A(0)\partial_sA(0)=\frac{1}{2\rho_m}(\tilde A-A)^2
  \pm  iq \int_0^{2\rho_m} ds'~\partial_{s'}(2 c(s'))  \left[A  \left( 1-\frac{s'}{2\rho_{\rm m}}\right) + \frac{s'}{2\rho_{m}}\tilde A\right]^2+\ldots$$

The matrix of current-current correlators is
$$
\left[\begin{matrix} <jj> & <j\tilde j>\cr <\tilde j j>&<\tilde j\tilde j>\cr \end{matrix}\right]
=\frac{N_3N_7}{2\pi^2}
\left[\begin{matrix} \frac{1}{\rho_m}\pm iq k_1 & - \frac{1}{\rho_m}\pm iqk_2\cr
 - \frac{1}{\rho_m}\pm iqk_2& \frac{1}{\rho_m}\pm iqk_3\cr \end{matrix}\right]
$$
where
$$
k_1=2\int_0^{2\rho_m} ds'~\partial_{s'}(2 c(s'))   \left( 1-\frac{s'}{2\rho_{\rm m}}\right)^2
$$
$$
k_2=2\int_0^{2\rho_m} ds'~\partial_{s'}(2 c(s'))  \left( 1-\frac{s'}{2\rho_{\rm m}}\right) \frac{s'}{2\rho_{m}}
$$
$$
k_3=2\int_0^{2\rho_m} ds'~\partial_{s'}(2 c(s'))  \left(  \frac{s'}{2\rho_{m}} \right)^2
$$
To linear order in $q$, the eigenvalues are
$$
<j_+j_+>=\pm iq\frac{N_3N_7}{2\pi^2}\frac{(k_1+2k_2+k_3)}{2}+\ldots
$$
$$
<j_-j_->=\frac{2}{\rho_m}\pm iq\frac{N_3N_7}{2\pi^2}\frac{(k_1-2k_2+k_3)}{2}+\ldots
$$
or
$$
<j_+j_+>=\mp iq\frac{N_3N_7}{4\pi } +\ldots
$$
$$
<j_-j_->=\frac{N_3N_7}{2\pi^2}\frac{2}{\rho_m}\pm iq\frac{N_3N_7}{2\pi^2}\int_0^{2\rho_m} ds'~\partial_{s'}(2 c(s'))
\left(1-2  \frac{s'}{2\rho_{m}} \right)^2+\ldots
$$
 which are the results quoted in eq.~(\ref{jminus}) and (\ref{jplus}). Furthermore, in this approximation,
 $j_\pm= j\pm\tilde j+{\cal O}(q)$ and the elementary charge quanta for $j_+$ and $j_-$ are identical to those for $j$
 and $\tilde j$.  By convention, we have taken these charges to be integers.

\end{document}